\documentclass[12pt,a4paper]{article}
\usepackage{amssymb}
\usepackage{epsf}
\usepackage{cite}
\newlength{\HFPP}       \HFPP5.4mm
\makeatletter
\@addtoreset{equation}{section}
\makeatother

\def\preprint#1#2{\noindent\hbox{#1}\hfill\hbox{#2}\vskip 10pt}

\newcommand{\one}[1]{\mathop{#1}\limits^1}
\newcommand{\two}[1]{\mathop{#1}\limits^2}
\newcommand{\be}[1]{\begin{equation}\label{#1}}
\newcommand{\ba}[1]{\begin{eqnarray}\label{#1}}
\newcommand{\ee}{\end{equation}}
\newcommand{\ea}{\end{eqnarray}}
\newcommand{\dis}{\displaystyle}
\newcommand{\non}{\nonumber\\\rule{0pt}{18pt}}
\newcommand{\nona}[1]{\nonumber\\\rule{0pt}{#1pt}}
\newcommand{\eq}[1]{(\ref{#1})}

\begin{document}
\begin{titlepage}
\def\thefootnote{\fnsymbol{footnote}}

\preprint{ITP-UH-24/98}{October 1998}
\vfill

\begin{center}
  {\Large\sc New solutions to the Reflection Equation and
	the projecting method}
\vfill

{\sc Holger Frahm}$^{\dagger}$
	and
{\sc Nikita A.\ Slavnov}$^{\ddag}$
\vspace{1.0em}

$^{\dagger}${\sl
  Institut f\"ur Theoretische Physik, Universit\"at Hannover\\
  D-30167~Hannover, Germany}\\
$^{\ddag}${\sl 
  Steklov Mathematical Institute, Moscow 117966, Russia}
\end{center}
\vfill
\begin{quote}
New integrable boundary conditions for integrable quantum systems can
be constructed by tuning of scattering phases due to reflection at a
boundary and an adjacent impurity and subsequent projection onto
sub-spaces.  We illustrate this mechanism by considering a
$gl(m<n)$-impurity attached to an open $gl(n)$-invariant quantum chain
and a Kondo spin $S$ coupled to the supersymmetric $t$--$J$ model.
\end{quote}

{Short title: }
	New solutions to the reflection equation

{PACS-Nos.:
72.10.Fk	
75.10.Jm	
75.30.Hx	
}
\vspace*{\fill}
\setcounter{footnote}{0}
\end{titlepage}

\section{Introduction}

Studies of integrable models with open boundary conditions have
attracted much interest recently.  The exact solutions of these
systems provide important insights into the nature of bound states due
to the presence of local potentials and properties of impurities
coupled to one-dimensional quantum systems
\cite{kask:96,essl:96,befr:97,frzv:97b,wang:97,hschulz:86,gohs:90,%
Wang97,Huxx98}.

The classification of open boundary conditions for integrable quantum
chains is possible within the framework of the quantum inverse
scattering method (QISM) \cite{vladb} by supplementing the Yang-Baxter
equation ---which guarantees the factorizability of $N$-particle
scattering processes in the bulk of the system--- with the reflection
equation (RE) algebra to ensure compatibility of two-particle
scattering and particle-boundary scattering \cite{cher:84,skly:88}.
The simplest solutions to this RE algebra are $c$-number matrices with
entries corresponding to the phase shifts due to (static) boundary
fields in the different channels.
In general, such boundary fields will break the symmetry of the model,
in spin chains they have been identified as magnetic fields acting on
the boundary sites \cite{skly:88}, for the $gl(2|1)$-invariant
(supersymmetric) $t$--$J$ model the (diagonal) $c$-number solutions of
the RE correspond to boundary chemical potential and boundary magnetic
fields, respectively \cite{gonz:94,essl:96}.
Dynamic impurities located at the boundary can also be described in
terms of solutions to the RE: as observed in Ref.~\cite{skly:88}
'dressing' of $c$-number boundary matrices with local monodromy
matrices generates new solutions to the RE with elements acting non
trivially in an impurity Hilbert space.  Such operator valued
solutions to the RE ---called `regular' in the following--- have been
used to construct models of spin-$S$ chains with spin-$S'$ impurities
located on the boundary site (see e.g.\ \cite{frzv:97b,wang:97}).
All of these models are similar in that operators acting on the
quantum space of the impurity need to be chosen among representations
of the \emph{same} algebra as the ones acting on the bulk sites, e.g.\
$SU(2)$ for Heisenberg models or $gl(2|1)$ for the supersymmetric
$t$--$J$ model, just as in the corresponding closed chain systems
\cite{anjo:84}.

Integrable models of Kondo impurities in one-dimensional electronic
continuum \cite{hschulz:86,gohs:90} (recently rediscovered in
Ref.~\cite{wavo:96}) and lattice models \cite{Wang97,Huxx98} which
have been solved by means of the coordinate Bethe Ansatz appear not to
fit into this scheme: In these systems the quantum space of the
impurity is a projection of the symmetry group onto a subgroup acting
only on the spin-degree of freedom.
Recently, Zhou and coworkers \cite{Zhou98.1,Zhou98.2} have succeeded
in formulating the model of a Kondo impurity in the
$gl(2|1)$-symmetric $t$--$J$ model \cite{Wang97,Huxx98} in the
framework of the RE algebra.  They have found an operator valued
solution to the RE which apparently cannot be obtained by the `regular'
dressing procedure with $gl(2|1)$-symmetric monodromy matrices
containing the impurity degrees of freedom.  Instead, they propose a
decomposition into `singular' matrices with $SU(2)$ spin operators as
entries.

In this paper, we introduce a method which allows projection of
`regular' solutions of the RE to a certain subspace of the impurity's
Hilbert space after adjusting the boundary phase shifts of the
$c$-number matrix to the ones due to the dressing impurity.  In the
following section we briefly review the RE formalism and formulate the
necessary conditions for the application of the projection method.  In
Section~\ref{sectgln} we apply this method to the case of $gl(n)$
algebra. Finally, we show how to obtain the `singular' boundary
matrices of Refs.~\cite{Zhou98.1,Zhou98.2} within this approach.

\section{General method}
Before consideration of the specific cases we would like 
to formulate our approach in general.

The classification of integrable boundary conditions within the QISM
is based on representations of two algebras ${\cal T}_\pm$
\cite{skly:88}.  The RE for ${\cal T}_-(u)$ has the form:
\begin{equation}
\label{REc}
R_{12}(u_1-u_2){\one{\cal T}}_-(u_1) R_{21}(u_1+u_2){\two{\cal
T}}_-(u_2) ={\two{\cal T}}_-(u_2)R_{12}(u_1+u_2) {\one{\cal
T}}_-(u_1)R_{21}(u_1-u_2)\ .
\end{equation}
Here we use standard notations: ${\one{\cal T}}_-(u) ={\cal
T}_-(u)\otimes I$ and ${\two{\cal T}}_-(u) =I\otimes{\cal T}_-(u)$.
The RE for ${\cal T}_+(u)$ will not be considered in the present
paper: the solutions of these equations are related to (\ref{REc}) by
an isomorphism, in the Hamiltonian limit ${\cal T}_\pm (u)$
determine the right and left boundary of the quantum chain,
respectively \cite{skly:88}.

The $R$-matrix satisfies quantum Yang--Baxter equation (YBE)
\be{QYBE}
R_{12}(u)R_{13}(u+v)R_{23}(v)=
R_{23}(v)R_{13}(u+v)R_{12}(u).
\ee
As usual $R_{21}(u)=P_{12}R_{12}(u)P_{12}$, where $P_{12}$ 
is the permutation operator. The unitarity property 
of the $R$-matrix assumed to be hold
\be{1unit}
R_{21}(u)R_{21}(-u)=\rho(u),
\ee
where $\rho(u)$ is a scalar function.

As we have mentioned already in the Introduction, operator-valued
(quantum) solution of the RE \eq{RE} can be constructed following
Ref.~\cite{skly:88}: let $L(u)$ be a quantum solution of the
intertwining equation of the Quantum Inverse Scattering Method:
\be{YBint}
R_{12}(u_1-u_2)\one{L}(u_1)\two{L}(u_2)=
\two{L}(u_2)\one{L}(u_1)R_{12}(u_1-u_2).
\ee
The entries of the $L$-operator are quantum operators, acting
in a Hilbert space $\cal H$.

Given a solution of (\ref{YBint}) we define an operator-valued matrix
$K_-(u)$ as
\be{factor}
K_-(u)=L(u){\cal T}(u)L^{-1}(-u) 
\ee
where ${\cal T}(u)$ is a $c$-number solution of \eq{REc}.  Then one
can check \cite{skly:88} that the quantum $K_-(u)$ boundary matrix
solves the RE: 
\be{RE}
   R_{12}(u_1-u_2){\one{K}}_-(u_1)
  R_{21}(u_1+u_2){\two{K}}_-(u_2)= {\two{K}}_-(u_2)R_{12}(u_1+u_2)
  {\one{K}}_-(u_1)R_{21}(u_1-u_2).  
\ee 
In what follows we shall refer to the formula \eq{factor} as `regular'
factorization.  Similarly, we call the corresponding $K$-matrix
`regular' solution of the RE.

In the paper \cite{Zhou98.2} a new type of the RE solution
had been found. This new $K$-matrix can not be presented in the
form \eq{factor}. Instead, the authors propose so-called `singular'
factorization
\be{singfactor}
K_-(u)\equiv K_{s}(u)=\lim\limits_{\epsilon\to 0}
L_\epsilon(u)L_\epsilon^{-1}(-u),
\ee
where $L_\epsilon$-operator depends on auxiliary parameter
$\epsilon$. The special feature of this solution is that factorization
\eq{singfactor} is valid for arbitrary $\epsilon$ (i.e.\ $K_{s}(u)$
does not depend on $\epsilon$) which allows to omit the limit in
\eq{singfactor}.  On the other hand, the operator $L_\epsilon$
satisfies the intertwining relation \eq{YBint} in the limit
$\epsilon\to0$ only, but the limit $\epsilon\to0$  for
$L_\epsilon^{-1}(u)$ does not exist. Following the
authors of \cite{Zhou98.2} we call the representation \eq{singfactor}
`singular' factorization and the corresponding $K$-matrix `singular'
solution of the RE, in spite of its well defined limit for
$\epsilon\to0$.

In the present paper we show that these `singular' solutions are
nothing but projections of suitably chosen `regular' ones.  Our
approach is based on the following simple observation.  Consider some
`regular' solution of the RE, obtained by the standard procedure
\eq{factor}.  The entries of such quantum $K$-matrix are operators,
acting in the same space ${\cal H}$, as the entries of the
$L$-operator.  Now consider two orthogonal sub-spaces ${\cal H}_1$ and
${\cal H}_2$, such that ${\cal H}_1 \oplus{\cal H}_2={\cal H}$,
characterized by projectors $\pi_1$ and $\pi_2$ respectively.  Then it
is easily seen that vanishing of one of the projections
$\pi_1K_-(u)\pi_2$ or $\pi_2K_-(u)\pi_1$
\be{orthogonal}
  \pi_1K_-(u)\pi_2=0,\qquad
  \mbox{or}\qquad \pi_2K_-(u)\pi_1=0,
\ee
implies that the projections $\pi_1K_-(u)\pi_1$ and $\pi_2K_-(u)\pi_2$ of the
operator $K_-(u)$ onto the sub-spaces ${\cal H}_1$ and ${\cal H}_2$ solve the
RE:
\ba{REproject}
&&\hspace{-21mm}{\dis
R_{12}(u_1-u_2)\left(\one{\pi_iK_-(u_1)\pi_i}\right)
R_{21}(u_1+u_2)\left(\two{\pi_iK_-(u_2)\pi_i}\right)}\non
&&\hspace{22mm}{\dis=
\left(\two{\pi_iK_-(u_2)\pi_i}\right)R_{12}(u_1+u_2)
\left(\one{\pi_iK_-(u_1)\pi_i}\right)R_{21}(u_1-u_2),}
\ea
where $i=1,2$.
Thus, new quantum solutions of the RE can be generated via projecting
of the original $K$-matrix onto a sub-space of its quantum Hilbert
space.

The first problem, however, is to find the decomposition ${\cal
H}_1\oplus{\cal H}_2={\cal H}$, possessing the property
\eq{orthogonal}.  For arbitrary $K_-(u)$ boundary matrix such a
decomposition may not exist.  Nevertheless, as will be demonstrated
below, this decomposition is possible for certain solutions of the RE
of the type (\ref{factor}) where the $c$-number factor has been
properly adjusted to the dressing $L$-operators.  In particular, the
solution of the RE found in Ref.~\cite{Zhou98.2}, just can be obtained
by the method described above.

The second problem related to this method, is whether the projecting
provides us with {\it really} new solutions of the RE, i.e.\ ones not
allowing `regular' factorization.  It is easy to see that it is not
always so.  If, for example, ${\cal H}_2$ is one-dimensional
sub-space, then evidently the projection $\pi_2K_-(u)\pi_2$ is just
one of the known $c$-number solutions of the RE.

Apart from this trivial possibility, the examples considered below do
not permit to formulate a criterion, which would allow one to predict
that a projection of a `regular' solution is not `regular'.  However,
we shall demonstrate, that `singular' solutions can be obtained via
projecting procedure.

%
\section{The case of $gl(n)$ algebra\label{sectgln}}
In this section we demonstrate the method of projecting,
using the example of $gl(n)$ algebra. Consider $n^2\times n^2$
$R$-matrix
\be{Rgln}
R(u)=uI+P,
\ee
where the permutation operator $P$ has the entries 
$P_{jk}^{\alpha\beta}=\delta_{j\beta}\delta_{k\alpha}$. The simplest
quantum $L$-operator, satisfying the equation \eq{YBint} has the form
\be{Loperatorgln}
L_{ij}(u)=\frac{1}{u+1}\Bigl(\delta_{ij}u+|j\rangle\langle i|\Bigr).
\ee
Here 
\be{defvectors}
\langle i|=(\underbrace{0,\dots,0}_{i-1},1,0\dots,0),
\qquad |i\rangle =(\langle i|)^T.
\ee
In fact, this $L$-operator coincides with the $R$-matrix \eq{Rgln}
up to normalization factor. The entries of the $L$-operator
act in the quantum space ${\cal H}=C^n$.

Introduce two quantum projectors $\pi_1$ and $\pi_2$:
\be{projec}
\pi_1=\sum_{k=1}^{m}|k\rangle\langle k|,\qquad
\pi_2=\sum_{k=m+1}^{n}|k\rangle\langle k|,\qquad
\pi_1+\pi_2=I_q,
\ee
where $m$ is a fixed number from the interval $1\le m\le n$, and
$I_q$ is identity operator in $\cal H$. Obviously, these projectors 
define two orthogonal sub-spaces:  ${\cal H}_1=\mathop{\rm 
span}\{|1\rangle,\dots,|m\rangle\}$ and ${\cal H}_2=\mathop{\rm 
span}\{|m+1\rangle,\dots,|n\rangle\}$.  At the first stage we are 
going to construct a `regular' $K$-matrix by means of the $L$-operator 
\eq{Loperatorgln} and some $c$-number solution of the RE. Then we 
shall consider the projections of this $K$-matrix onto sub-spaces 
${\cal H}_1$ and ${\cal H}_2$.

We start with the solution of the RE 
\be{startsol} 
 K_-(u)=L(u+c){\cal T}(u)L^{-1}(-u+c).  
\ee 
Here $c$ is a constant, ${\cal T}(u)$ is a diagonal $c$-number
solution of the RE breaking the $gl(n)$-symmetry of the system down to
$gl(m)$ \cite{vego:93}:
\be{c-numsol}
{\cal T}_{ij}(u)=\delta_{ij}h_i(u).
\ee
Here 
\be{hi}
 h_i(u)=1\quad \mbox{for}\quad i\le m;\qquad
 h_i(u)\equiv 
 h(u)=\frac{\xi-u}{\xi+u}\quad \mbox{for}\quad i>m\ ,
\ee 
with some constant $\xi$.

With the normalization in (\ref{Loperatorgln}) we have 
$L^{-1}(-u)=L(u)$. Thus we arrive at $K_-(u)=K_d(u)+K_a(u)$, where
\begin{eqnarray}
  (K_d(u))_{ij}&=&\frac{\delta_{ij}}{(u+1)^2-c^2}
  \left[(u^2-c^2)h_i(u)+\sum_{k=1}^{n}
  h_k(u)|k\rangle\langle k|\right],
\nonumber\\
\label{Kxij}\\
  (K_a(u))_{ij}&=&\frac1{(u+1)^2-c^2}
  \Bigl[(u+c)h_i(u)+(u-c)h_j(u)\Bigr]
  |j\rangle\langle i|.
\nonumber
\end{eqnarray}
Thus, the `regular' solution of the RE \eq{RE} is constructed. Next 
let us consider the projections of this solution.  First, we have to 
adjust the parameters in (\ref{Kxij}) such that $\pi_1K_-(u)\pi_2=0$
or $\pi_2K_-(u)\pi_1=0$.  The projections of the part $K_d(u)$
automatically are equal to zero
\be{d1122}
\pi_1K_d(u)\pi_2=\pi_2K_d(u)\pi_1=0.
\ee
As for the projections of the part $K_a(u)$, we have
\be{a12}
\Bigl(\pi_1K_a(u)\pi_2\Bigr)_{ij}=\left\{
\begin{array}{cll}
\Bigl(K_a(u)\Bigr)_{ij},&\qquad&  i>m,\quad j\le m;\non
0,&\qquad&\mbox{otherwise};
\end{array}\right.
\ee
\be{a21}
\Bigl(\pi_2K_a(u)\pi_1\Bigr)_{ij}=\left\{
\begin{array}{cll}
\Bigl(K_a(u)\Bigr)_{ij},&\qquad&  i\le m,\quad j> m;\non
0,&\qquad&\mbox{otherwise}.
\end{array}\right.
\ee
Thus, by choosing $\xi=\pm c$ in \eq{hi} we obtain
$\pi_1K_-(u)\pi_2=0$ ($\pi_2K_-(u)\pi_1=0$). In either cases the
projections $\pi_1K_-(u)\pi_1$ and $\pi_2K_-(u)\pi_2$ satisfy the RE.
We would like to emphasize especially that the parameter $\xi$ in the
$c$-number solution (\ref{c-numsol}) has to be adjusted to the
parameter $c$ in the dressing $L$-operators for the projections
$\pi_1K_-(u)\pi_2$ and $\pi_2K_-(u)\pi_1$ to vanish.

Let us now focus on $\xi=c$:  in this case the projected reflection
matrices are
%
\begin{eqnarray}
  \Bigl(\pi_1K_-(u)\pi_1\Bigr)_{ij}&=&
  \left\{
  \begin{array}{cl}
  \dis
  \frac{(u^2-c^2+1)\delta_{ij}+2u|j\rangle\langle i|}
  {(u+1)^2-c^2},&\qquad 
  i,j\le m;  \nona{26}
  \dis
  \delta_{ij}\frac{c+1-u}{c+1+u}&\qquad\mbox{otherwise};
  \end{array}\right.
\nonumber\\
\label{proj11+c}\\
  \Bigl(\pi_2K_-(u)\pi_2\Bigr)_{ij}&=&\frac{c-u}{c+u}
  \left\{
  \begin{array}{cl}
  \dis
  \frac{(u^2-c^2+1)\delta_{ij}+2u|j\rangle\langle i|}
  {(u+1)^2-c^2},&\qquad 
  i,j> m;  \nona{26}
  \dis
  \delta_{ij}\frac{c-1+u}{c-1-u}&\qquad\mbox{otherwise}.
  \end{array}\right.
\nonumber
\end{eqnarray}
Introducing $L$-operators, acting in the sub-spaces ${\cal H}_1$ and
${\cal H}_2$ only:
\begin{eqnarray}
  \Bigl(L_1(u)\Bigr)_{ij}&=& (u+c)\delta_{ij} + |j\rangle\langle i|,
  \qquad i,j\le m,
\nonumber\\
\label{L12}\\
  \Bigl(L_2(u)\Bigr)_{ij}&=& (u-c)\delta_{ij} + |j\rangle\langle i|,
  \qquad i,j> m.
\nonumber
\end{eqnarray}
the projections (\ref{proj11+c}) can be presented as block-matrices
\begin{eqnarray}
  \pi_1K_-(u)\pi_1&=&\frac{c+1-u}{c+1+u}
  \left(
  \begin{array}{cc}
  L_1(u)L_1^{-1}(-u)&0\non
  0&1
  \end{array}\right),
\nonumber\\
\label{proj11+c1}\\
  \pi_2K_-(u)\pi_2&=&\frac{c-u}{c+u}\,\,\frac{c-1+u}{c-1-u}
  \left(
  \begin{array}{cc}
  1&0\non
  0&L_2(u)L_2^{-1}(-u)
  \end{array}\right).
\nonumber
\end{eqnarray}
Clearly, the external factors can be removed, and we arrive at two new
solutions of the RE
\begin{equation}
  K_{s1}(u)=\left(
  \begin{array}{cc}
  L_1(u)L_1^{-1}(-u)&0\non
  0&1
  \end{array}\right),
\qquad
  K_{s2}(u)=\left(
  \begin{array}{cc}
  1&0\non
  0&L_2(u)L_2^{-1}(-u)
  \end{array}\right).
\label{Ks12}
\end{equation}
While these solutions can not be presented as \emph{regular} solutions
(\ref{factor}) of the RE they can be factorized in terms of
\emph{singular} solutions to (\ref{YBint}): with
\be{Lsing}
L_\epsilon (u)=
\left(\begin{array}{cc}
L_1(u)&0\non
0&\epsilon
\end{array}\right).
\ee
we can write $K_{s1}(u)=L_\epsilon(u) L_\epsilon(-u)^{-1}$.  However 
the operator $L_\epsilon(u)$ satisfies the equation \eq{YBint} only in the 
limit $\epsilon\to 0$. Thus, we have the complete analogy with the
case, considered in Ref.~\cite{Zhou98.2}.

In the conclusion of this section we would like to mention some 
properties of `singular' factorization, which make it essentially 
different from the `regular' one. First, inserting a $c$-number 
solution ${\cal T}(u)$ between dressing $L_\epsilon$-operators
\be{inserT}
K_{s,{\cal T}}=L_\epsilon(u){\cal T}(u)L_\epsilon^{-1}(-u),
\ee
we do not arrive at a new RE solution. The matrix \eq{inserT} does 
not satisfy the RE.  One should not be surprised of this fact, since, 
as we have seen, vanishing of projections $\pi_1K_-(u)\pi_2$ (or 
$\pi_2K_-(u)\pi_1$) was provided only due to the special choice
of the $\cal T$-matrix \eq{c-numsol}.

Second, in the `regular' case one can generate new $K$-matrices
via replacement
\[
L(u)\to T(u)=L_N(u)\cdots L_1(u),
\]
where $L_i(u)$ are copies of the original $L$-operator, acting in
different quantum spaces.  For the `singular' factors (\ref{Lsing})
this method fails, i.e. if $T_\epsilon= L_{\epsilon,N}\cdots
L_{\epsilon,1}$, then $K_{s,T}(u)=T_\epsilon(u)T^{-1}_\epsilon(-u)$
does not solve the RE.  This fact also can be explained in the
framework of the projecting method.  The matter is that the subtle
tuning of boundary and impurity properties which leads to the
fulfilment of the necessary condition (\ref{orthogonal}) cannot be
done in the large quantum space ${\cal H}_T$ of the matrices
$T_\epsilon(u)$ and $K_{s,T}(u)$.  This makes it impossible to find a
decomposition ${\cal H}_T={\cal H}_{T,1} \oplus{\cal H}_{T,2}$.


\section{Kondo impurity in the supersymmetric $t$--$J$ model}


Our second example deals with the Kondo impurity in the supersymmetric
$t$--$J$ model recently constructed in
Refs.~\cite{Wang97,Huxx98,Zhou98.1,Zhou98.2}.  Integrability of the
periodic model is proven by constructing of the enveloping vertex
model within a $\mathbf{Z}_2$-graded extension of the QISM
\cite{kul.86,esko:92,foka:93}.  A similar extension of the RE is
necessary, for the algebra ${\cal T}_-$ it is formally identical to
the ungraded case (\ref{RE})
with a $9\times9$ $R$-matrix
\be{Rmatr}
R_{12}(u)=uI+P_{12}\ .
\ee
Here $P_{12}$ is the $\mathbf{Z}_2$-graded permutation operator
\begin{equation}
\label{permut}
  \left(P_{12}\right)_{jk}^{\alpha\beta}=
	(-1)^{[j][\alpha]} \delta_{j\beta}\delta_{k\alpha}\ .
\end{equation}
The $\mathbf{Z}_2$-grading is chosen in such a way that $[1]=[2]=1$
and $[3]=0$.  The $R$-matrix \eq{Rmatr} satisfies the unitarity
property (\ref{1unit}).

The diagonal $c$-number solutions of the RE are again of the form
(\ref{c-numsol}) and correspond to boundary magnetic fields and
chemical potentials, respectively \cite{gonz:94,essl:96}.  Recently, a
new type of quantum solution of the RE \eq{RE} has been found
\cite{Zhou98.2}:
\be{chinasol}
K_s(u)=\left(
\begin{array}{ccc}
\alpha(u)+\beta(u)S^z,&\beta(u)S^-&0\non
\beta(u)S^+&\alpha(u)-\beta(u)S^z,&0\non
0&0&1
\end{array}\right).
\ee
Here $S^z$, $S^{\pm}$ are usual generators of a $SU(2)$ 
algebra: $[S^z,S^{\pm}]=\pm S^{\pm}$,\ $[S^{+},S^{-}]
=2S^z$,\ $\mathbf{S}^2=s(s+1)$. The functions $\alpha(u)$ and
$\beta(u)$ are equal to
\begin{eqnarray}
\alpha(u)&=&\frac{(c+s+1/2)(c-s-1/2)-u^2+u}
{(c+s+1/2-u)(c-s-1/2-u)},
\nonumber\\
\label{alpha}\\
\beta(u)&=&\frac{2u}{(c+s+1/2-u)(c-s-1/2-u)},
\nonumber
\end{eqnarray}
with a constant $c$.

The general structure of the $K$-matrix \eq{chinasol} looks very 
similar to \eq{proj11+c}, \eq{Ks12}. Indeed, this solution can be
presented in terms of `singular' factorization \cite{Zhou98.2}:
\be{singfact}
K_s(u)=L_\epsilon(u){L_\epsilon}^{-1}(-u),
\ee
where
\be{singL}
L_\epsilon(u)=\left(
\begin{array}{ccc}
\dis u-c-1-S^z&\dis -S^-&\dis 0\non
\dis-S^+&\dis u-c-1+S^z&\dis 0\non
\dis 0&\dis 0&\dis\epsilon
\end{array}\right).
\ee
Just as in our previous example, the operator $L_\epsilon$ satisfies
the (graded version of the) intertwining equation \eq{YBint} in the
limit $\epsilon\to0$ only.  All the `pathological' properties of the
`singular' solutions, listed in the end of the previous section, are
valid for the $K$-matrix \eq{chinasol}.  This leads us to assume that
in fact the $K$-matrix \eq{chinasol} is nothing but a projection of a
`regular' solution of the RE.

To reproduce the result \eq{chinasol} of Ref.~\cite{Zhou98.2} by means
of the projecting method we have to consider solutions of the
intertwining relation (\ref{YBint}) and the reflection equation
(\ref{RE}) invariant under the action of the graded Lie algebra
$gl(2|1)$ (see e.g.\ \cite{snr:77,marcu:80}).  Apart from the
generators $1$, $S^z$, $S^{\pm}$ forming an (ungraded) $gl(2)$
subalgebra it has an additional generator $B$ of even parity (charge),
commuting with the spin operators, and four odd generators $V^\pm$ and
$W^\pm$.  The commutation relations between even and odd generators
are listed below
\be{commoddeven}
\begin{array}{c}
[S^z,V^\pm]=\pm\frac12 V^\pm,\qquad
[S^\pm,V^\pm]=0,\qquad
[S^\mp,V^\pm]=V^\mp,\non
[S^z,W^\pm]=\pm\frac12W^\pm,\qquad
[S^\pm,W^\pm]=0,\qquad
[S^\mp,W^\pm]=W^\mp,\non
[B,V_\pm]=\frac12V_\pm,\qquad
[B,W_\pm]=-\frac12W_\pm.
\end{array}
\ee
The odd generators satisfy anticommutation relations
\be{anticommodd}
\begin{array}{c}
\{V^\pm,V^\pm\}=\{V^\pm,V^\mp\}=
\{W^\pm,W^\pm\}=\{V^\pm,W^\mp\}=0,\non
\{V^\pm,W^\pm\}=\pm\frac12S^\pm,\qquad
\{V^\pm,W^\mp\}=\frac12(S^z\pm B).
\end{array}
\ee
In the following we shall consider the `atypical' representation
$[s]_+$ of this algebra \cite{snr:77,marcu:80}.  In a basis
$\{|b,s,m\rangle\}$ where $B$, $\mathbf{S}^2$ and $S^z$ are diagonal
this representation contains two spin multiplets of spin $s$ and
$s-1/2$ with charge $b=s$ and $s+1/2$, respectively:
\begin{equation}
\label{H1H2} 
	{\cal H}_1=\mathop{\rm span}\{|s,s,m\rangle\},
 \qquad {\cal H}_2=\mathop{\rm span}\{|s+1/2,s-1/2,m\rangle\}\ .
\end{equation}
The nonvanishing matrix elements of the remaining operators are
\begin{eqnarray}
&&\langle s+\frac12,s-\frac12,m\pm\frac12|S^\pm
  |s+\frac12,s-\frac12,m\mp\frac12\rangle=\sqrt{s^2- m^2}
\nonumber\\
&&\langle s+\frac12,s-\frac12,m\pm\frac12|V^\pm|s,s,m\rangle=
  \pm\sqrt{\frac{s\mp m}{2}}
\label{matrelements}\\
&&\langle s,s,m|W^\pm
  |s+\frac12,s-\frac12,m\mp\frac12\rangle
	=\sqrt{{s\pm m}\over{2}}
\nonumber
\end{eqnarray}

Now we consider the following `regular' quantum solution of the RE
\be{2solutionRE} 
  K_-(u)=L(u+c){\cal T}(u)L^{-1}(-u+c).  
\ee
Here ${\cal T}(u)$ is the $c$-number solution of the RE corresponding
to a boundary chemical potential:
\be{cnumbersol}
{\cal T}(u)=\mathop{\rm diag}
\left(1,1,\frac{\xi-u}{\xi+u}\right) 
\ee
and the $L$-operator of the containing the degrees of freedom of the
quantum impurity in \eq{2solutionRE} is equal to \cite{kul.86}
\be{Loperator}
(u-s-1/2)L(u)=u-s-1/2+\left(
\begin{array}{ccc}
\dis B-S^z&\dis -S^-&\dis -\sqrt2V^-\non
\dis -S^+&\dis B+S^z&\dis \sqrt2V^+\non
\sqrt2W^+&\dis \sqrt2W^-&\dis 2B
\end{array}
\right).  
\ee 
We have chosen the normalization such that $L^{-1}(-u+c)=L(u-c)$.

For the projection of the `regular' $K$-matrix (\ref{2solutionRE}) we
use the decomposition of the impurity quantum space $\cal H$ into
direct sum ${\cal H}={\cal H}_1\oplus{\cal H}_2$ of spaces
(\ref{H1H2}).  The projection of \eq{2solutionRE} onto sub-spaces
${\cal H}_1$ and ${\cal H}_2$ it easily done by computing the
projections of the $L$-operator and to use
\be{sumprojections}
\pi_iK_-\pi_j= [\pi_iL\pi_1]{\cal T}[\pi_1L^{-1}\pi_j]+
[\pi_iL\pi_2]{\cal T}[\pi_2L^{-1}\pi_j], 
\ee
where $\pi_i$ are projectors onto ${\cal H}_i$, as before.  These
calculations are quite straightforward, therefore we summarize the
results only:
The condition $\pi_1K_-(u)\pi_2=0$ is satisfied by choosing
$\xi=c+s-1/2$ in \eq{cnumbersol}.  Then the projection
$\pi_1K_-(u)\pi_1$ exactly coincides with the matrix $K_s(u)$
\eq{chinasol}.  Thus, as we have stated above, the `singular' RE
solution of Ref.~\cite{Zhou98.2} is indeed the projection of the
`regular' solution (\ref{2solutionRE}).

\section{Conclusion}
We have presented a method which allows ---by adjusting the parameters
of the $c$-number boundary matrix and those of an adjacent dynamical
impurity--- to construct new quantum solutions of the RE by means of
the projection (\ref{REproject}).  Since $K_-(u)$ is directly related
to the boundary term of the corresponding quantum hamiltonian
\cite{skly:88} satisfying the condition (\ref{orthogonal}) amounts
to (block--) diagonalization of the hamiltonian in the Hilbert space
of the impurity.  While each of these blocks may correspond to a
previously known boundary condition ---as trivially seen when projecting
to a one-dimensional subspace--- we have presented several cases where
new representations of the RE algebra arise which do not allow to be
presented in terms of `regular' factorization (\ref{factor}).  These
new cases include models for a $gl(m<n)$-spin impurity coupled to a
$gl(n)$-symmetric quantum chain and the case of an $SU(2)$ Kondo-spin
in the supersymmetric $t$--$J$ chain \cite{Zhou98.1,Zhou98.2}.
A common feature of these `singular' solutions to the RE is a 
remaining non-trivial symmetry in the impurity degrees of freedom
\emph{after} projection.

The existence of projected boundary matrices has important consequences
for the solution of systems with open boundary conditions by means of the
algebraic Bethe ansatz: proper choice of a suitable reference state, which
needs to be contained in the projected Hilbert space, is crucial to capture
the properties of the impurity site.  This statement holds in particular
for the graded models such as the $t$--$J$ model where \emph{different}
Bethe ans\"atze are possible starting from various fully polarized states.

Finally we would like to emphasize the remark of Ref.~\cite{Zhou98.2}
regarding Kondo-impurities in \emph{closed} chains: it is obvious from
the discussion above that the presence of a boundary next to the quantum
impurity is essential for our construction.  Using a `singular' 
$L$-operator such as $L_{\epsilon\to0}(u)$ from (\ref{singL}) to 
construct a periodic chain leads to the Heisenberg model with 
impurity of Andrei and Johannesson \cite{anjo:84} rather than a Kondo 
spin in a $t$--$J$ model.


\section*{Acknowledgments}
This work has been supported by the Deutsche Forschungsgemeinschaft
under Grant No.\ Fr~737/2--3.

\end{document}